\documentstyle[aps,twocolumn,epsf,rotate]{revtex}
\begin{document}
\draft
\twocolumn[\hsize\textwidth\columnwidth\hsize\csname
@twocolumnfalse\endcsname

\title{
The Attractive Hubbard Model in 2D: Is it capable of describing a
pseudogap and preformed pairs?}

\author{M. Letz}
%\footnote{Present address: Institut f\"ur Physik,
%Johannes-Gutenberg Universit\"at, 55099 Mainz, Germany} 
%and R. J. Gooding}

%\address{Dept. of Physics, Queen's University, Kingston, ON Canada K7L
%3N6\\
%$^*$ Present address: 
\address{Institut f\"ur Physik,
Johannes-Gutenberg Universit\"at, 55099 Mainz, Germany}

\date{\today}

\maketitle

\begin{abstract}

Deviations from Fermi liquid behavior are well documented in the 
normal state of the cuprate superconductors, and some of these differences
seem to be related to pre-transitional features appearing at
temperatures above T$_c$. The observation of a pseudogap, e.g. in ARPES 
experiments, is a familiar example of this physics. One potential explanation 
for this behaviour involves preformed pairs with
finite lifetimes existing in the normal state above T$_c$. In this
way two characteristic temperatures can be established. A higher one
T$^*$ at which pairs begin to form and the actual T$_c$ at which a
phase-coherent superconducting phase is established.

In order to test these ideas we have investigated the negative U Hubbard
model in two dimensions in the fully self-consistent ladder
approximation at low electron densities.
In the non self-consistent version of this
theory the system always shows an instability towards Bose-condensation of
infinite lifetime pairs. In contrast to this, pairs obtain a finite lifetime
due to pair-pair interaction and the sharp two-particle bound state
is strongly lifetime broadened when self-consistency is applied.
A quasi-particle scattering rate which varies linearly with temperature
is also found.

The fully self-consistent calculation we were able to perform using
a ${\bf {\vec k}}$--averaged approximation in which the self-energy
loses its ${\bf {\vec k}}$-dispersion due to a ${\bf {\vec
k}}$-average. This approximation is found to preserve the essential physics.


\end{abstract}
]
%\vskip -0.5cm
%\twocolumn
\narrowtext

\section{Introduction:}
\label{sec:intro}

In three dimensions (3D) the negative U Hubbard model has two well
understood borderline cases: One is the weak coupling limit $U \ll
t$ (t is the transfer). In this case the model shows superconductivity
of mean-field, 
BCS type with a large coherence length of the cooper pairs. The second
well understood case is the limit of strong coupling {\it and} low
densities $U \gg t$, $ n \approx 0$. In this case pairs of electrons
which now form composite Bosons condense into a two particle bound
state and the Fermi surface is destroyed.

In 2D however the situation is different. Here it has been
shown by Schmitt-Rink et al. \cite{svr} that for any coupling
strength  the Fermi surface is 
lost and Bose condensation (at T=0) takes place. The calculation that
was used in \cite{svr} was a T-matrix approximation (taking into
account ladder diagrams in the particle--particle channel) in its non
self-consistent, non conserving form which is only valid in the zero
density limit, $n \rightarrow 0$.

The motivation to study a 2D system with low but finite densities and
intermediate coupling strength ($U \approx$ bandwidth $W$) comes from
the high-T$_c$ cuprates. In these systems the normal state transport is
governed by 2D CuO$_2$ planes, the quasiparticle density is low but
finite (one still finds a Fermi surface) and the coherence length of a
cooper pair is small ($\xi \approx $3-4 lattice constants). 
However the negative U Hubbard model in its simplest form shows only
s-wave pairing. But we believe 
that full understanding of the simple s-wave problem is a necessary
condition to extend the calculations to more complicated models.
We are confident that many parts of the
physics developed here for the s-wave case will survive when more
complicated d-wave pairing is considered.

\section{Calculation}

When trying to expand the known results for the zero density limit
towards finite densities the main additional interaction which has to
be taken into account is the interaction between pairs. The lowest
order interaction term would be the exchange of two
electrons between two pairs. Such interactions are included into the
equations by extending the non selfconsistent work \cite{svr} to a
fully selfconsistent as has been discussed in
e.g. \cite{haussmann93,fresard92}. In order to perform such a fully
selfconsistent calculation we use dynamical mean field theory
\cite{infdrev} which in our particular problem becomes not only exact
in infinite dimensions but also for the limit of large correlations
($U > t$)
which is in our particular problem even more important.

In this case the {\bf k} independent self-energy (we denote the
k-average with over-lined quantities) 
\begin{equation}
\label{eq:sigscf}
\overline{\Sigma}(i \omega _n) =  \frac{1}{\beta} \sum_{m}
\overline{\Gamma} (i \omega _m + 
i \omega _n) G(i \omega _m) 
\end{equation}
results from a {\bf K} independent vertex function $\overline{\Gamma}(i
\Omega_n)$. The vertex function now consists of two different parts:
The one particle continuum and the bound state. In the large U limit
the {\bf K}-dispersion of the bound state vanishes and therefore the
{\bf K} averaged vertex function will already be a good approximation 
in two dimensions. To calculate this vertex function from the
susceptibility we apply another approximation
\begin{equation}
\overline{\Gamma}(i \Omega )
\approx \frac{U^2 \overline{\chi}
(i\Omega_n)  }{(1 - U  \overline{\chi}
(i\Omega_n) ) } 
\end{equation}
By doing this the next term which is neglected is of the order of the mean
square deviation of the susceptibility, $\overline{\chi}^2 -
\overline{\chi^2}$ as is discussed in detail in \cite{kaveragelang}.
To get the full set of equations to solve selfconsistently we further
need the equation for $\chi$
\begin{equation}
\label{eq:chi}
\chi({\bf K},i \Omega _n) = -\frac{1}{N \beta} \sum_{m,{\bf k}}  
G({\bf K}-{\bf k},i \Omega _n - i \omega _m) G({\bf k},i \omega _m) 
\end{equation}
and for the one particle Green function:
\begin{equation}
G({\bf k},i \omega _n) = \left ( G^0({\bf k},i \omega _n)^{-1} - 
\overline{\Sigma}(i
\omega _n) \right ) ^{-1} \label{eq:greenscf}~~.
\end{equation}
When solving these equations non selfconsistently we reproduce the
results of \cite{svr}. That means the Fermi surface is lost at low
temperatures and we get Bose condensation into the two--particle bound
state.

\section{Results and Discussion}

\begin{figure}
\unitlength1cm
\epsfxsize=10cm
\begin{picture}(7,7.5)
\put(-2.2,-0.3){\rotate[r]{\epsffile{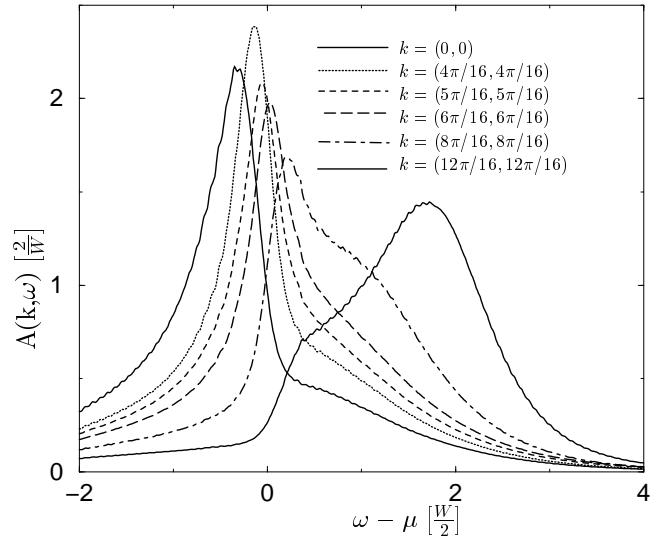}}}
\end{picture}
\caption{For a density of n = 0.3, an attractive correlation of U = -8t,
and a temperature of k$_B$T = 0.1t, the k-dispersion
of the one particle spectral function is shown. The self-energy was
obtained by a fully selfconsistent T-matrix calculation and the
k-points are choosen along the (1,1) direction. Especially at the
Fermi energy we obtain a quasiparticle peak whereas the incoherent
broadening comes mainly into play away from k$_F$. The energy units
along the axis are chosen in units of half the bandwidth $\left [
\frac{W}{2} \right ] $ }
\label{fig:ako}
\end{figure}

When doing a fully selfconsistent calculation the situation changes
drastically; the infinite lifetime bound state gets strongly lifetime
broadened and merges with the one particle continuum. We further regain a
Fermi surface \cite{kaveragelang}. In Fig.\ref{fig:ako} we have
plotted the dispersion of the one particle density of states A({\bf
k},$\omega$) where the 
self-energy was obtained via a fully selfconsistent calculation on a
quasi-2D system with a constant initial density of states. We have
chosen several k-points along the (1,1) direction.
The correlation strongly broadens and renormalizes the one particle
peak but at the Fermi energy we obtain a clear quasiparticle peak. We
therefore find {\it no} pseudo gap from our calculation.

From our results it seems to follow that at low but finite
quasiparticle densities the intuitive picture of pairs of electrons
which condense as composite Bosons breaks down. In the following we
discuss some arguments which support this result:
When thinking in such a Bose picture the
effective hopping of pairs is given by second order perturbation
theory. The kinetic energy of such a pair is therefore given by an
effective hopping of $\frac{t^2}{U}$. But only in the zero density
limit the interaction between such Bosons can be considered to be
small. At finite densities the dominant interaction is given by the
fact that a pair has a smaller number of virtual hopping processes due
to the existence of other pairs. Such interaction is therefore caused
by the Pauli principle, is repulsive can not be neglected in
comparison with the kinetic 
energy of the pair. Such kind of discussion is well known in nuclear
physics \cite{otsuka78}. When one therefore maps the Hamiltonian with
strongly interacting Fermions onto Bosons one ends up, at finite
density, with strongly interacting Bosons which does not solve the
problem.

\acknowledgements

The author wishes to thank R.~J.~Gooding, F.~Marsiglio and A.~Chernyshev
for intense discussions.
This work was supported by the ''Deutsche Forschungsgemeinschaft"
and by the NSERC of Canada. 

% \begin{thebibliography}{10}
%\bibliographystyle{unsrt}
%\bibliography{lit}

\end{document}